# ARTICLE

# Polar and apolar light-induced alignment of ferroelectric nematic at photosensitive polymer substrate

Ruslan Kravchuk,[ab] Oleksandr Kurochkin,[ab] Vassili G. Nazarenko,[*ab] Volodymyr Sashuk,[ab] Mykola Kravets,[ab] Bijaya Basnet [cd] and Oleg D. Lavrentovich [*cde]

Surface alignment of a recently discovered ferroelectric nematic liquid crystal ($N_F$) is usually achieved by buffed polymer films, which produce a unidirectional polar alignment of the spontaneous electric polarization. We demonstrate that photosensitive polymer substrates could provide a broader variety of the alignment modes. Namely, a polyvinyl cinnamate polymer film irradiated by linearly polarized ultraviolet (UV) light yields two modes of surface orientation of the $N_F$ polarization: (1) a planar apolar mode, in which the equilibrium $N_F$ polarization aligns perpendicularly to the polarization of normally impinging UV light; the $N_F$ polarization adopts either of the two antiparallel states; (2) a planar polar mode, produced by an additional irradiation with obliquely impinging UV light; in this mode, there is only one stable azimuthal direction of polarization in the plane of the substrate. The two modes differ in their response to an electric field. In the planar apolar mode, the polarization can be switched back and forth between two states of equal surface energy. In the planar polar mode, the field-perturbed polarization relaxes back to the single photoinduced "easy axis" once the field is switched off. The versality of modes and absence of mechanical contact makes the photoalignment of the $N_F$ attractive for practical applications.

## INTRODUCTION

Alignment of nematic (N) liquid crystals (LCs) is crucial for their applications. While historically the first alignment technique employed mechanical rubbing of glass substrates or of alignment layers such as polymers,[1-3] an exceedingly popular modern technique is that of photoalignment, in which a photosensitive substrate is exposed to light with a pre-designed polarization, incident angle, doze, and intensity.[4-11] Photoalignment allows one to produce complex patterns of molecular orientation with high spatial resolution,[12,13] controllable strength of surface anchoring,[14,15] dynamic alignment[16] and realignment,[17,18] and the capability to pattern molecular orientations on flexible and curved substrates.[11,19] Photoalignment avoids mechanical contact with the substrate and thus does not induce impurities, electric charges, or mechanical damage to the treated surfaces, unlike conventional rubbing. Under irradiation a substrate coated with a photoalignment material becomes anisotropic and produces an "easy axis" of nematic alignment, orientation of which depends the polarization and incident angle of light, mechanism of light-induced modifications of the substrate, and the nature of the N material.[9-11] There are three types of photosensitive transformations used to align an N: (i) photoisomerization of compounds such as azobenzene derivatives, (ii) photo-crosslinking and (iii) photodegradation of polymers.[9-11] In most cases, the photoinduced easy axis is perpendicular to the polarization of light.[9-11]

In the literature cited above, photoalignment has been successful in the alignment of the director $\hat{\mathbf{n}}$ of a conventional paraelectric N with the property $\hat{\mathbf{n}} \equiv -\hat{\mathbf{n}}$.[9-11] The prevailing types of alignment are the so-called "planar", in which $\hat{\mathbf{n}}$ makes a small "pretilt" angle $\psi$ with the substrate, $\psi = 1° - 6°$, and a tilted alignment, with a higher $\psi$. Because of the identity $\hat{\mathbf{n}} \equiv -\hat{\mathbf{n}}$, pretilted planar alignment produced by normally incident linearly polarized light is two-fold degenerate and requires an additional step such as oblique irradiation to lift the degeneracy and produce a single "easy axis" of director orientation.[9]

An intriguing question is whether the photoalignment approach can be applied to the recently discovered ferroelectric nematic ($N_F$)[20-23] with polar orientation of molecules carrying large dipole moments, which yields spontaneous macroscopic electric polarization $\mathbf{P}$. Spontaneous polarization imposes polar symmetry onto the material properties since the states $\mathbf{P}$ and $-\mathbf{P}$ are not equivalent. In the materials explored so far, the vector $\mathbf{P}$ is collinear with the director $\hat{\mathbf{n}} \equiv -\hat{\mathbf{n}}$ that specifies the average orientation of the long axes of the $N_F$ molecules. The presence of electric polarization $\mathbf{P}$ makes orientations with a pretilt $\psi \neq 0$ at dielectric substrates questionable since a substrate-piercing polarization deposits a strong surface

[a.] Institute of Physics, National Academy of Sciences of Ukraine, Prospect Nauky 46, Kyiv, 03028, Ukraine. E-mail: vnazaren@iop.kiev.ua
[b.] Institute of Physical Chemistry, Polish Academy of Sciences, Kasprzaka 44/52, 01-224 Warsaw, Poland
[c.] Advanced Materials and Liquid Crystal Institute, Kent State University, Kent, OH 44242, USA. E-mail: olavrent@kent.edu
[d.] Materials Science Graduate Program, Kent State University, Kent, OH 44242, USA
[e.] Department of Physics, Kent State University, Kent, OH 44242, USA
† Electronic Supplementary Information available. See DOI: 10.1039/x0xx00000x

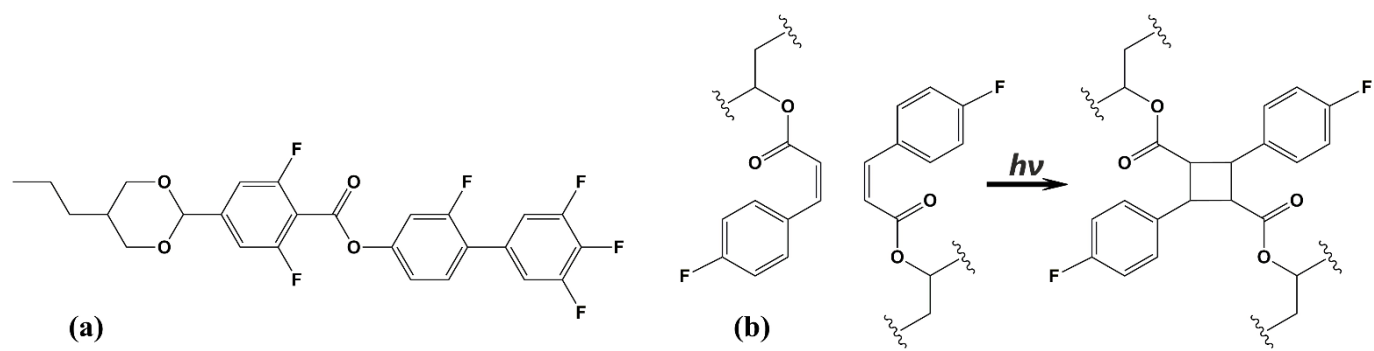

**Fig. 1** (a) Chemical structure of the ferroelectric nematic liquid crystal DIO and (b) PVCN-F and its photochemical transformation.

charge, as studies of mechanically rubbed substrates[24-31] or interfaces with isotropic media suggest.[32-35] A spectacular illustration of avoidance of the normal component of polarization at interfaces with dielectric materials is that in the curved channels with $N_F$, **P** aligns everywhere tangentially to the spatially-varying walls.[36] Nevertheless, there are reports that a homeotropic alignment at the $N_F$-air can be achieved for an $N_F$ material doped with an ionic polymer.[37] Furthermore, since $\mathbf{P} \neq -\mathbf{P}$, a unidirectional buffing of a polymer aligning layer produces structural polarity of the aligning layer; molecular interactions translate this surface polarity into the bulk polarization.[24,26,29,38,39] For example, the **P** of the $N_F$ phase in the material abbreviated DIO aligns antiparallel to the buffing direction on polyimide substrates,[26,29,40] while **P** of RM734 aligns parallel[41] to this direction.

As compared to alignment by buffing, photoalignment of the $N_F$ is much less explored. So far, only photoalignment by azodyes Brilliant Yellow[42,43] and SD1[44,45] have been explored; in all cases, a normally incident linearly polarized light has been utilized. The mechanism of alignment produced by azodyes is well known[9-11,46] and involves a trans-cis photoisomerization of the azobenzene moieties. Under a linearly polarized ultraviolet (UV) irradiation, the trans-cis and cis-trans isomerization continues till the transition dipole of the molecules becomes orthogonal to the polarization of light. This approach produces two-fold degenerate planar alignment, as the states with **P** and $-\mathbf{P}$ perpendicular to the light polarization are of equal probability to form.[42-45] Besides this two-fold degeneracy, another disadvantage of the alignment by photoisomerization cited in the literature[9-11] is its instability under heat and light. In fact, our research revealed that Brilliant Yellow does not align well the high-temperature $N_F$ phase of material RM734 at temperatures 100 °C and above.

In comparison, photo-crosslinking materials undergoing irreversible photochemical reactions that limit molecular movements show a better prospect for applications.[9,11] Popular materials of this type are cinnamate-based polymers.[6,7,47] In this work, we explore one of these materials, polyvinyl-4(fluoro-cinnamate) (PVCN-F), introduced by Gerus' and Reznikov's group,[48-54] as a photoalignment layer for the $N_F$ material DIO. DIO, first synthesized by Nishikawa et al.,[21] is formed by fluorinated rod-like molecules with strong longitudinal dipole moments. The goal of the study is to explore whether the photoalignment technique can produce unidirectional polar alignment of **P** as opposed to the two-fold degenerate planar alignment. So far, there is no experimental evidence that such a polar orientation could be created by photoalignment, although such a possibility is expected on the ground of symmetry.[38] We explore two geometries of irradiation: (1) normal incidence of linearly polarized light and (2) oblique incidence of linearly polarized light. Linearly polarized UV irradiation of PVCN-F induces in-plane anisotropy through a covalent molecular cross-linking and an easy axis perpendicular to the polarization of light. We demonstrate that the linearly-polarized UV-irradiated PVCN-F substrates yield apolar and polar modes of surface alignment depending on the irradiation geometry. In the apolar mode, normally incident light irradiation produces two-fold degenerate alignment, in which the $N_F$ polarization **P** can be switched by an in-plane electric field between two states of the same surface energy. In the polar mode, additional irradiation with oblique incidence yields only one azimuthal direction of the polarization **P**. If **P** is realigned by an external electric field, this polar easy axis of alignment is restored once the field is removed. The polar photoalignment mode of the $N_F$ represents a significant technological advance since this non-contact process avoids problems such as buffing-induced electric charges and impurities, to which the $N_F$ is much more sensitive than its paraelectric N counterpart.

## MATERIALS AND IRRADIATION PROTOCOLS

We explore the $N_F$ material 2,3',4',5'-tetrafluorobiphenyl-4yl 2,6-difluoro-4-(5-propyl-1,3-dioxan-2-yl) benzoate known as DIO,[21] Fig. 1a. On cooling from the isotropic (I) phase, the phase sequence of DIO, synthesized as described by Brown et al.[55] is I – 174°C – N – 82°C – SmZ$_A$ – 66°C – $N_F$ – 34°C – Crystal, where SmZ$_A$ is an antiferroelectric smectic phase.[56]

PVCN-F of molecular weight 30,000 has been synthesized by Dr. I. Gerus in the Institute of Bioorganic Chemistry and Petrochemistry, Kyiv, Ukraine.[57] Prior to UV irradiation, the fluorinated cinnamoyl groups of PVCN-F show a small out-of-plane preference in the alignment, most likely caused by the fluorine atom that imparts hydrophobic properties onto these groups.[57] Normally incident linearly polarized light realigns the

Journal Name ARTICLE

cinnamoyl groups perpendicularly to the polarization direction **L** of light.[57]

Flat sandwich cells are formed by two glass plates with transparent indium-tin-oxide (ITO) electrodes and spin-coated PVCN-F layers as described by Bugayova et al.[52] Each glass plate with the PVCN-F coating is photoaligned separately before the assembly, by 60 s exposure to the normally impinging linearly polarized UV light from the broadband spectrum source Panacol UV P-280. The UV intensity at the polymer surface is fixed at 10 mW/cm$^2$. Light irradiation of cinnamate-containing polymers causes photodimerization of the cinnamoyl fragments through breaking of the unsaturated C=C bonds and their replacement with the saturated ones, Fig.1b. As a result of this photoinduced transformation, the number of cinnamoyl groups aligned along the polarization direction **L** is reduced, while their number oriented perpendicularly to **L** increases.[7,54,57,58] As noted by Ichimura,[59] the cinnamate groups experience also trans-cis isomerization and align perpendicularly to **L** because of it, similarly to azodyes. Exploration of the particular case of PVCN-F demonstrated that both photoeffects induce the anisotropy[54] that aligns polar N molecules such as 4′-pentyl-4-cyanobiphenyl (5CB) along the direction perpendicular to **L**.[48,51-53,57,59]

We use two UV-irradiation protocols, Fig. 2. In the protocol 1, the linearly polarized UV light beam is normal to the substrate; the light polarization $\mathbf{L}_1$ is along the $y$-axis. The LC cells in this case are formed by two irradiated PVCN-F coating facing each other with the directions $\mathbf{L}_1$ being parallel to each other. In the photoalignment protocol 2, the normal UV irradiation of step 1 of a duration 60 s is followed by 300s irradiation with an obliquely incident UV light with polarization $\mathbf{L}_2$ in the $xz$ plane; the wave vector $\mathbf{k}_2$ is in the $xz$ plane and makes an angle 45$^0$ with the substrate normal, Fig. 2. The cells are assembled with two plates for which the projections of $\mathbf{k}_2$ onto the substrate are parallel to each other. The LC slabs are of a thickness (5.0-5.4) μm. Two stripes of ITO electrodes are located on one glass plate to apply an in-plane electric field, Fig. 2.

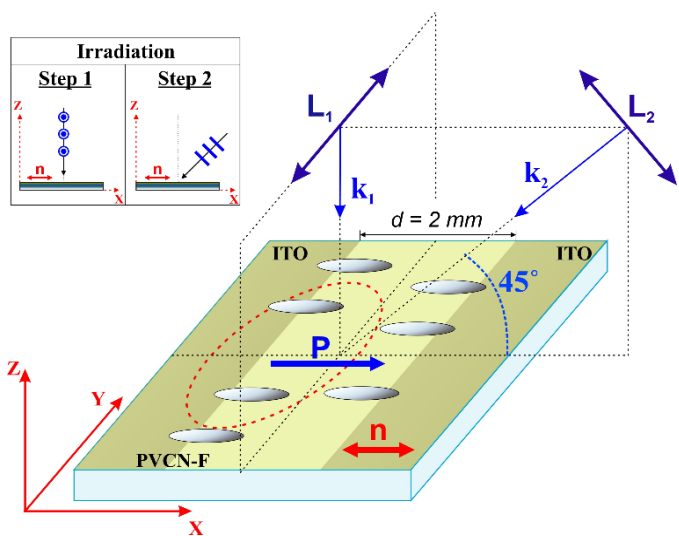

**Fig. 2** Schemes of light irradiation of PVCN-F substrate. The ITO strip electrodes at a glass substrate are separated by 2 mm to produce a uniform in-plane field. In the photoalignment mode 1 the wave vector $\mathbf{k}_1$ of the linearly polarized UV light beam is normal to the substrate; the light polarization $\mathbf{L}_1$ is along the $y$-axis. In the photoalignment mode 2, the normal UV irradiation of mode 1 is followed by irradiation with an obliquely incident UV light with linear polarization $\mathbf{L}_2$ in the $xz$ plane; the wave vector $\mathbf{k}_2$ is in the $xz$ plane and makes an angle 45° with the substrate normal. The red dashed ellipse encloses the typical area of optical microscopy observation.

## RESULTS AND DISCUSSION

### Apolar photoalignment mode

Irradiation protocol 1 with 60 s of normally impingent UV light with linear polarization, $\mathbf{L} = (0, \pm 1, 0)$, yields a homogeneous alignment of the N phase, with the director $\hat{\mathbf{n}} = (\pm 1, 0, 0)$ perpendicular to **L**. Between two crossed polarizers, the texture is dark when $\hat{\mathbf{n}}$ is along the polarizer or analyzer, Fig. 3a. Cooling down the sample into the $N_F$ phase results in a texture with multiple stripe domains elongated along the $y$-axis, Fig. 3b. The central parts of the domains are dark but the domain walls separating the domains are not extinct. The textures observed with the analyzer rotated counterclockwise from the polarizer, Fig. 3c, and clockwise, Fig. 3d, differ little from each other, and do not produce optical contrast between the neighbouring domains, which indicates that the optical axis and the polarization **P** in the domains do not twist along the normal $z$ to the cell. The light transmittance through the domains measured as a function of the angle $\gamma$ between the polarizer and analyzer shows a minimum at $\gamma = 0$, Fig. 3e. All these observations suggest that the polarization **P** is along the $x$-axis, but alternates in sign, from $\mathbf{P} = P(1,0,0)$ in one domain to $\mathbf{P} = P(-1,0,0)$ in the next domain. As explained previously,[43] the formation of uniform domains with antiparallel **P** is caused by the tendency of the material to reduce the depolarization field which would be significant in the case of a monocrystalline polar alignment of **P**. Alternating uniform stripe domains mitigate the depolarization effect when the slabs are relatively thin, a few microns.[43] An increase of the cell thickness produces twisted domains,[43,60] in which **P** twists around the normal $z$ to the cell. In our case, a small (less than 5%) number of such twisted domains is also observed, as verified by their optical activity in observations with uncrossed polarizers.

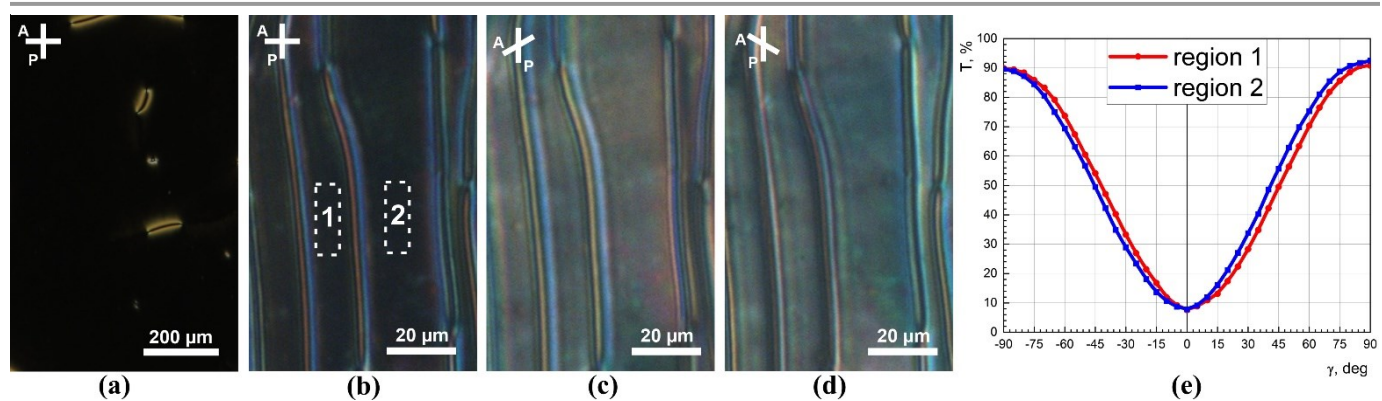

**Fig. 3** Apolar planar photoalignment by normally incident linearly polarized UV light. (a) A uniform monodomain textures of the N phase of DIO between two layers of PVCN-F observed between crossed polarizers, 95°C. (b) Polydomain $N_F$ texture, the polarization directions in adjacent domains are antiparallel to each other, 56°C. (c) and (d) The same, the analyzer is rotated counterclockwise and clockwise with respect to the analyzer by 40°, respectively. The textures are captured using a green interferometric filter with a centre wavelength λ = 531nm. (e) Light transmission through a small area in two domains marked by 1 and 2 in (b) as a function of the angle $\gamma$ between the analyzer and polarizer. Minimum transmittance at $\gamma = 0$ indicates that there is no twist along the normal $z$ to the cell.

The measured birefringence of DIO in homogeneous domains is $\Delta n \sim 0.2$ at the wavelength of light 532 nm, which is consistent with the previously measured values[26,29] indicating that there is no detectable pretilt angle in the $N_F$ phase and the orientation corresponds electrostatically required tangential alignment. The tangential alignment avoids a strong surface bound charge. Even a small tilt $\psi \sim 5^\circ$ of **P** from the $xy$ plane would produce a surface charge density $P_z \sim P\psi \sim 4 \times 10^{-3}\ \text{C m}^{-2}$, which is larger than the typical surface charge $(10^{-4} - 10^{-5})\ \text{C m}^{-2}$ of adsorbed ions reported for nematics;[61,62] here $P \approx 4.4 \times 10^{-2}\ \text{C m}^{-2}$ is the polarization of DIO.[21]

One concludes that protocol 1 of photoalignment with normally incident UV light produces uniformly aligned domain of polarization with apolar surface anchoring along both directions of the $x$-axis normal to the polarization of light: the polarity of **P** in neighboring domains alternates from $\mathbf{P} = P(1,0,0)$ to $\mathbf{P} = P(-1,0,0)$. The domain structure can be erased by a relatively weak in-plane direct current (DC) electric filed $\mathbf{E} = E(\pm1,0,0)$, where $E \sim 100$ V/m, applied along the $x$-axis. The field realigns polarization direction in the domains with **P** antiparallel to **E**; the domains with **P** along **E** do not respond to the field. Upon reversing the field polarity, the response is reversed.

Homogeneous monodomain alignment of **P** can be achieved by cooling the photoaligned sample from the N phase to the $N_F$ phase in the presence of an applied DC electric field $\mathbf{E}_{\text{cooling}} = 100$ V/m. The uniform alignment forms in the gap between the ITO electrodes but also at the surface of the electrodes; it is preserved when the field is switched-off, Fig. 4a, Video S1. Repeated application of the field along the polar direction of **P** does not change the texture much, Fig. 4b. The reversal of field polarity realigns **P** by $180^0$ in the gap between the ITO stripes;

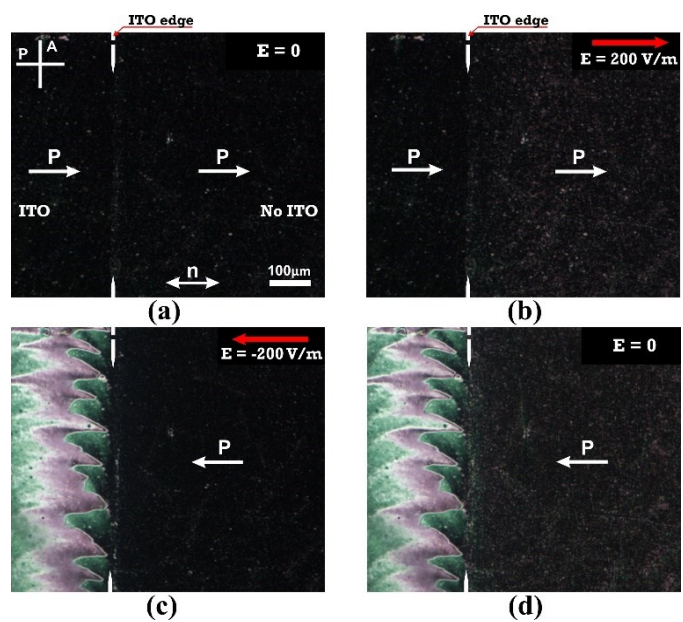

**Fig. 4** Apolar photoalignment assisted by the DC electric field. **(a)** Initial uniform texture of $N_F$ after cooling in electric field $\mathbf{E}_{\text{cooling}}$. Polarizations **P** (white arrows) is aligned uniformly in the gap between the ITO stripes and at the ITO surfaces. (b) The $N_F$ texture remains uniform when the electric field $\mathbf{E} \uparrow\uparrow \mathbf{P}$ is applied again in the same direction as during the cooling process. (c) The $N_F$ texture after field polarity reversal. The polarization in the gap between the electrodes is uniform but realigned by $180^0$. At the ITO surfaces, the structure is misaligned. (d) The uniform $N_F$ texture persists after the field is switched off.

the texture above the ITO becomes misaligned, with domains of apparent twists of opposite handedness, separated by zigzag domain walls, Fig. 4c. The field required for reversal of **P** is $\mathbf{E}_{\text{reversal}} \approx 200$ V/m. Upon removal of the field $\mathbf{E}_{\text{reversal}}$, the alignment remains stable for 2-3 hours, Fig. 4d, after which time the uniform alignment in the gap gradually degrades, apparently because of the frustration between the memory effect of the $\mathbf{E}_{\text{reversal}}$ and $\mathbf{E}_{\text{cooling}}$ fields.

The polarization realignment from $\mathbf{P} = (P,0,0)$ to $\mathbf{P} = (-P,0,0)$ and back requires electric fields $\sim 100$ V/m for both directions of switching; the difference does not exceed 10 V/m. This demonstrates that the UV-irradiated PVCN-F coated surfaces produce a strong quadrupolar in-plane $N_F$ alignment; the polar contribution cannot be detected. This quadrupolar alignment mode is drastically different from the unipolar $N_F$ alignment induced by polymer substrate with a substantial polar component.[26]

**Polar photoalignment**

The symmetry of the described bistable photoalignment is broken when the UV irradiation is performed in two steps, according to the protocol 2, in which the initial normal incidence of the UV beam is followed by irradiation with an oblique incidence, Fig. 2. This second irradiation is performed with linearly polarized UV light of intensity $I$ = 10 mW/cm² for $t$ = 300 s. The polarization direction is perpendicular to the light polarization in the first step, Fig. 2. The angle of incidence is 45°.

The cells are assembled in a parallel fashion. The material is cooled from the N to the $N_F$ phase in the absence of any electric field. The resulting alignment is uniform with the $N_F$ director perpendicular to the polarization $\mathbf{L}_1$ of normally incident irradiation. Alignment of $\mathbf{P}$ is polar, along the direction that is opposite to the projection of wavevector $\mathbf{k}_2$ onto the substrate, i.e., along the $x$ axis in Fig. 2: $\mathbf{P} = P(1,0,0)$, Fig. 5a. Application of the DC electric field stabilizes the $N_F$ texture within the domains when $\mathbf{E} = E(1,0,0)$ is along $\mathbf{P}$, Fig. 5b. The opposite direction of $\mathbf{E} = E(-1,0,0)$ realigns $\mathbf{P}$, Fig. 5c, Video S2. The cell with the two-step irradiation demonstrates monostable ground state. After removal of the electric field $\mathbf{E} = E(-1,0,0)$, $\mathbf{P}$ relaxes to the initial state $\mathbf{P} = (P,0,0)$, within (10-200) s, Fig. 5d. The single equilibrium direction of $\mathbf{P}$ indicates that oblique irradiation induces a unipolar easy axis.

Qualitatively similar polar alignment is observed when the second stage of irradiation lasts last than 300 s or when the substrates are irradiated only with an obliquely incident UV light, linearly polarized or not polarized at all. However, the quality of alignment in these modes is worse than in the protocol 2 with two sequential irradiation steps

The polar in-plane anchoring in $N_F$ cells is described by the anchoring potential

$$W(\varphi) = \tfrac{1}{2} W_Q \sin^2\varphi - W_P(\cos\varphi - 1) \quad (1)$$

with the two energy minima, one global at $\varphi = 0$, and another local at $\varphi = \pm\pi$. [26,38] Here, $W_Q \geq 0$ and $W_P \geq 0$ are the apolar (quadrupolar, or nematic-like) and polar anchoring coefficients, respectively, $\varphi$ is the angle between $\mathbf{P}$ and the $x$ axis. When $W_P = 0$, the anchoring is polarity-insensitive, and the minima at $\varphi = 0; \pi$ are equal. This case corresponds to the protocol 1 with a single normal irradiation of PVCN-F substrate, Fig. 3. The oblique irradiation produces a nonzero $W_P$, thus the alignment at $\varphi = 0$ is energetically preferable than that at $\varphi = \pi$. Since the field-assisted alignment with $\varphi = \pi$ spontaneously relaxes into $\varphi = 0$, this state does not correspond even to a local minimum of $W(\varphi)$, which means that $W_P \geq W_Q$ in the protocol 2 of the photoalignment, Fig. 5. The polar component of photoinduced alignment is very strong as compared to the case of buffing-induced anchoring at a polymer substrate in Ref. [26], in which case $W_P \approx 0.1 W_Q$.

## Conclusions

We demonstrate two different photoalignment modes of $N_F$ at a photosensitive polyvinyl cinnamate polymer substrate.

(1) Planar apolar mode 1, in which the normally irradiated PVCN-F surface provides a strong quadrupolar in-plane anchoring of the polarization $\mathbf{P}$. $\mathbf{P}$ aligns perpendicularly to the polarization of normally impinging UV light. The direction of $\mathbf{P}$ can be switched back and forth between two collinear states of opposite polarity of the same surface energy.

(2) Planar polar mode 2, produced by an additional oblique irradiation with polarized UV light that results in only one equilibrium in-plane alignment of $\mathbf{P}$. Oblique irradiation even with unpolarized beam creates a director tilt away from the substrate in photoalignment of the N phase. [9] The reason is that photosensitive substrates develop anisotropy along the direction perpendicular to the polarization of light. For unpolarized obliquely incident light there is only one such anisotropy direction, which is the tilted direction of the wavevector of light. The obliquely irradiated PVCN-F substrate thus tends to align the nematic director $\hat{\mathbf{n}}$ in a tilted fashion, along the wavevector $\mathbf{k}_2$ of light. However, polar character of the $N_F$ ordering requires $\mathbf{P}$ to be tangential to any interface with a dielectric medium to avoid deposition of bound charge. As a result, the combination of photoalignment-preferred tilt at the photosensitive substrate, electrostatically required tangential alignment of the $N_F$, and different affinity of the heads and tails of the polar $N_F$ molecules to the substrate produce a polar in-plane alignment of $\mathbf{P}$, either along the projection of $\mathbf{k}_2$ onto the substrate or antiparallel to it. The alignment polarity of $\mathbf{P}$ should depend on the affinity of tails and heads of the polar molecules to the substrate. Experimental data, Fig. 2, suggest that the negatively charged fluorinated ends of the DIO molecules show a better affinity to the PVCN-F than the aliphatic tails, which justifies the alignment of $\mathbf{P}$ along the direction antiparallel to the projection of $\mathbf{k}_2$ onto the substrate. The details of nanoscale surface arrangements of $N_F$ molecules require further studies.

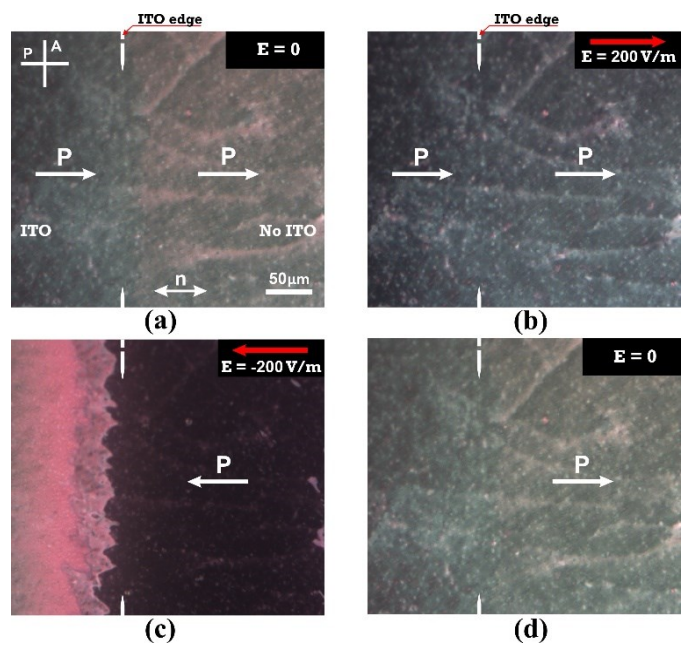

**Fig. 5** Mode 2 of photoalignment. (a) A uniform texture of DIO at PVCN-F coated surfaces observed between crossed polarizers in the $N_F$ phase; 56°C. (b) The electric field is applied along $\mathbf{P}$. The polarization $\mathbf{P}$ within the interelectrode region and outside it coincides. (c) The texture of the $N_F$ phase under the opposite electric field, $\mathbf{E} \uparrow\downarrow \mathbf{P}$. $\mathbf{P}$ reorients along the applied field $\mathbf{E}$. (d) The initial texture is restored when the electric field is removed (E = 0). The dashed line indicates the edge of the electrode.

Note that the polar anchoring induced in the protocol 2 is rather strong, with the corresponding anchoring coefficient $W_P$ in Eq.1 being larger than the quadrupolar counterpart $W_Q$. For comparison, a unidirectional buffing of polyimide substrates produces a noticeably lower values, $W_P \approx 0.1 W_Q$, Ref. [26]. The results offer an intriguing possibility to vary the ratio $W_P/W_Q$ in a broad range.

Both reported schemes of photoalignment of the ferroelectric nematic liquid crystal avoid electrostatic problems often associated with mechanical treatments such as polymer buffing and offer flexibility in the design of different orientational patterns of spontaneous electric polarization with polar and quadrupolar components. The polar alignment in mode 2, besides avoiding a mechanical contact with the substrates, yields a strong polar/quadrupolar ratio of anchoring coefficients. Both modes might be of a better utility as compared to the alignment by buffing.

## Author contributions

Ruslan Kravchuk: investigation. Oleksandr Kurochkin: investigation, writing – review & editing. Vassili G. Nazarenko: writing – original draft, supervision, resources, project administration, conceptualization. Volodymyr Sashuk: writing – review & editing, project administration. Mykola Kravets: investigation, chemical synthesis. Bijaya Basnet: investigation. Oleg D. Lavrentovich: writing – review & editing, project administration, conceptualization. The manuscript was written through contributions of all authors. All authors have given approval to the final version of the manuscript.

## Conflicts of interest

There are no conflicts to declare.

## Acknowledgements

This work was supported by NSF grant DMR-2341830 (O.D.L.), NASU project No. 0123U100832 (V.G.N., O.K., R.K.), and the Long-term program of support of the Ukrainian research teams at the Polish Academy of Sciences carried out in collaboration with the U.S. National Academy of Sciences with the financial support of external partners via the agreement No. PAN.BFB.S.BWZ.356.022.2023 (V.G.N., O.K., R.K., V.S., M.K.). The authors also acknowledge funding from the NATO SPS project G6030 (V.G.N., O.D.L.).

38  X. Chen, E. Korblova, M. A. Glaser, J. E. Maclennan, D. M. Walba and N. A. Clark, *Proc. Natl. Acad. Sci. U. S. A.*, 2021, **118**, e2104092118.
39  N. Sebastián, M. Čopič and A. Mertelj, *Phys. Rev. E*, 2022, **106**, 021001.
40  H. Kamifuji, K. Nakajima, Y. Tsukamoto, M. Ozaki and H. Kikuchi, *Appl. Phys. Express*, 2023, **16**, 071003.
41  B. Basnet, S. Paladugu, O. Kurochkin, O. Buluy, N. Aryasova, V. G. Nazarenko, S. V. Shiyanovskii and O. D. Lavrentovich, *Nat. Commun.*, 2025, **16**, 1444.
42  N. Sebastián, M. Lovšin, B. Berteloot, N. Osterman, A. Petelin, R. J. Mandle, S. Aya, M. J. Huang, I. Drevenšek-Olenik, K. Neyts and A. Mertelj, *Nat. Commun.*, 2023, **14**, 3029.
43  M. O. Lavrentovich, P. Kumari and O. D. Lavrentovich, *Nat. Commun.*, 2025, **16**, 6516.
44  J. T. Pan, B. H. Zhu, L. L. Ma, W. Chen, G. Y. Zhang, J. Tang, Y. Liu, Y. Wei, C. Zhang, Z. H. Zhu, W. G. Zhu, G. X. Li, Y. Q. Lu and N. A. Clark, *Nat. Commun.*, 2024, **15**, 8732.
45  C. Y. Li, X. Y. Xu, J. D. Yang, Y. Liu, L. Y. Sun, Z. J. Huang, S. Chakraborty, Y. Zhang, L. L. Ma, S. Aya, B. X. Li and Y. Q. Lu, *Sci. Adv.*, 2025, **11**, eadu7362.
46  G. J. Fang, J. E. Maclennan, Y. Yi, M. A. Glaser, M. Farrow, E. Korblova, D. M. Walba, T. E. Furtak and N. A. Clark, *Nat. Commun.*, 2013, **4**, 1521.
47  N. Klopcar, I. Drevensek-Olenik, M. Copic, M. W. Kim, A. Rastegar and T. Rasing, *Mol. Cryst. Liq. Cryst.*, 2001, **368**, 4163–4170.
48  D. Andrienko, Y. Kurioz, Y. Reznikov, C. Rosenblatt, R. Petschek, O. Lavrentovich and D. Subacius, *J. Appl. Phys.*, 1998, **83**, 50–55.
49  O. Buluy, Y. Reznikov, K. Slyusarenko, M. Nobili and V. Reshetnyak, *Opto-Electron. Rev.*, 2006, **14**, 293–297.
50  E. Ouskova, Y. Reznikov, S. V. Shiyanovskii, L. Su, J. L. West, O. V. Kuksenok, O. Francescangeli and F. Simoni, *Phys. Rev. E*, 2001, **64**, 051709.
51  I. Gerus, A. Glushchenko, S. B. Kwon, V. Reshetnyak and Y. Reznikov, *Liq. Cryst.*, 2001, **28**, 1709–1713.
52  L. Bugayova, I. Gerus, A. Glushchenko, A. Dyadyusha, Y. Kurioz, V. Reshetnyak, Y. Reznikov and J. West, *Liq. Cryst.*, 2002, **29**, 209–212.
53  S. Faetti, G. C. Mutinati and I. Gerus, *Mol. Cryst. Liq. Cryst.*, 2004, **421**, 81–93.
54  B. Sapich, J. Stumpe, I. Gerus and O. Yaroshchuk, *Mol. Cryst. Liq. Cryst.*, 2000, **352**, 443–452.
55  S. Brown, E. Cruickshank, J. M. D. Storey, C. T. Imrie, D. Pociecha, M. Majewska, A. Makal and E. Górecka, *ChemPhysChem*, 2021, **22**, 2506–2510.
56  X. Chen, V. Martinez, E. Korblova, G. Freychet, M. Zhernenkov, M. A. Glaser, C. Wang, C. H. Zhu, L. Radzihovsky, J. E. Maclennan, D. M. Walba and N. A. Clark, *Proc. Natl. Acad. Sci. U. S. A.*, 2023, **120**, e2217150120.
57  O. Yaroshchuk, T. Sergan, J. Kelly and I. Gerus, *Jpn. J. Appl. Phys.*, 2002, **41**, 275–279.
58  Y. Iimura, S. Kobayashi, T. Hashimoto, T. Sugiyama and K. Katoh, *IEICE Trans. Electron.*, 1996, **E79-C**, 1040–1046.
59  K. Ichimura, Y. Akita, H. Akiyama, K. Kudo and Y. Hayashi, *Macromolecules*, 1997, **30**, 903–911.
60  P. Kumari, B. Basnet, M. O. Lavrentovich and O. D. Lavrentovich, *Science*, 2024, **383**, 1364–1368.
61  R. N. Thurston, J. Cheng, R. B. Meyer and G. D. Boyd, *J. Appl. Phys.*, 1984, **56**, 263–272.
62  V. G. Nazarenko and O. D. Lavrentovich, *Phys. Rev. E*, 1994, **49**, R990–R993.